\documentclass[aps,prb,superscriptaddress,showpacs,reprint]{revtex4-2}

\usepackage[utf8]{inputenc}
\usepackage{amsmath,amsfonts,amssymb}
\usepackage{pifont}
\usepackage{rotating}
\usepackage{enumerate}
\usepackage{graphicx}    
\usepackage{epstopdf}
\usepackage{color}
\usepackage{soul} 
\usepackage{subfigure}
\usepackage{multirow}
\usepackage{bm}

\usepackage{hyperref}
\definecolor{dark-red}{rgb}{0.9,0.15,0.15}
\definecolor{dark-blue}{rgb}{0.15,0.15,0.4}
\definecolor{medium-blue}{rgb}{0,0,0.5}
\hypersetup{
    colorlinks={true},
     linkcolor={medium-blue},
    citecolor={blue},
     urlcolor={medium-blue}
}

\begin{document}
\title{CoRuVSi: A potential candidate for spin semimetal with promising spintronic and thermoelectric properties}
\author{Jadupati Nag}
\affiliation{Department of Physics, Indian Institute of Technology Bombay, Mumbai 400076, India}

\author{R. Venkatesh}
\affiliation{UGC-DAE Consortium for Scientific Research, University Campus, Khandwa Road, Indore-452001, India}

\author{Ajay Jha }
\affiliation{CRANN, AMBER, and School of Physics, Trinity College Dublin, Ireland}

\author{Plamen Stamenov}
\affiliation{CRANN, AMBER, and School of Physics, Trinity College Dublin, Ireland}

\author{P. D. Babu}
\affiliation{UGC-DAE Consortium for Scientific Research, Mumbai Centre, BARC Campus, Mumbai 400085, India}

\author{Aftab Alam}
\email{aftab@phy.iitb.ac.in}
\affiliation{Department of Physics, Indian Institute of Technology Bombay, Mumbai 400076, India}

\author{K. G. Suresh}
\email{suresh@phy.iitb.ac.in}
\affiliation{Department of Physics, Indian Institute of Technology Bombay, Mumbai 400076, India}


\begin{abstract}
Based on our experimental and theoretical studies, we report the identification of the quaternary Heusler alloy, CoRuVSi as a new member of the recently discovered spin semimetals class. Spin polarised semimetals possess a unique band structure in which one of the spin bands shows semimetallic nature, while the other shows semiconducting/insulating nature. Our findings show that CoRuVSi possesses interesting spintronic and thermoelectric properties. It crystallizes in perfect cubic structure with a partial L2$_1$-type disorder at room temperature. Magnetization data reveal a weak ferri-/antiferro magnetic ordering at low temperatures, with only a very small moment $\sim$ 0.13  $\mu_B$/f.u., attributed to the disorder. Transport results provide strong evidence of semimetallicity  dominated by two-band conduction, while magnetoresistance data show a non-saturating, linear, positive, magnetoresistance. Spin polarization measurements using point-contact Andreev reflection spectra reveal a reasonably high spin polarization of $\sim$ 50\%, which matches fairly well with the simulated result. Furthermore, CoRuVSi shows a high thermopower value of $0.7$ $m Watt/ m-K^{2}$ at room temperature with the dominant contribution from the semimetallic bands, rendering it as a promising thermoelectric material as well. Our ab-initio simulation not only confirms a unique semimetallic feature, but also reveals that the band structure hosts a linear band crossing at $\sim$ -0.4 eV below the Fermi level incorporated by a band-inversion. In addition, the observed topological non-trivial features of the band structure is corroborated with the simulated Berry curvature, intrinsic anomalous Hall conductivity and the Fermi surface. The partial L2$_1$ disorder is simulated using a special quasi random structure, which plays a crucial role in correctly explaining the magnetism and anomalous Hall effect. The simulated anomalous Hall conductivity for ordered and L2$_1$ disordered phase of CoRuVSi is found to be 102 and 52 S/cm, the later agrees fairly well with the experimentally measured value (45 S/cm). The coexistence of many interesting properties relevant for spintronic, topological and thermoelectric applications in a single material is extremely rare and hence this study could promote a similar strategy to identify other potential materials belonging to same class.
\end{abstract}

\date{\today}

\maketitle

\section{Introduction}
Heusler alloys are known to exhibit exotic phenomena as well as novel potential applications, which have stimulated a tremendous interest in physics and materials technology. Many systems from this family are reported to be promising spintronic materials such as half-metals (HM) \cite{PhysRevLett.50.2024} spin gapless semiconductors (SGS),\cite{bainsla2015spin} bipolar magnetic semiconductors (BMS),\cite{PhysRevB.104.134406} spin-valve \cite{PhysRevB.105.144409} etc. Most of the reported Heusler materials are superior to other materials from the application point of view because of their stable structure and high spin-polarization. The co-existence of different and interesting properties in these systems gives rise to new avenues for multifunctional materials suitable for technological applications such as  spintronics. Recently, tuning the electronic structure by defects/impurities has become a major focus by various researchers to achieve the desired properties suitable for applications.\cite{PhysRevB.103.085202} As Heusler alloys are prone to anti-site disorder, complex magnetic/electronic structures can be realized in these materials, with a wide tuning capability. One of the main motives of this work is to understand the role of anti-site disorder in band engineering and hence in the tuning of the magneto-electronic properties.

\begin{figure}[t]
 \centering
 \includegraphics[width=0.98\linewidth]{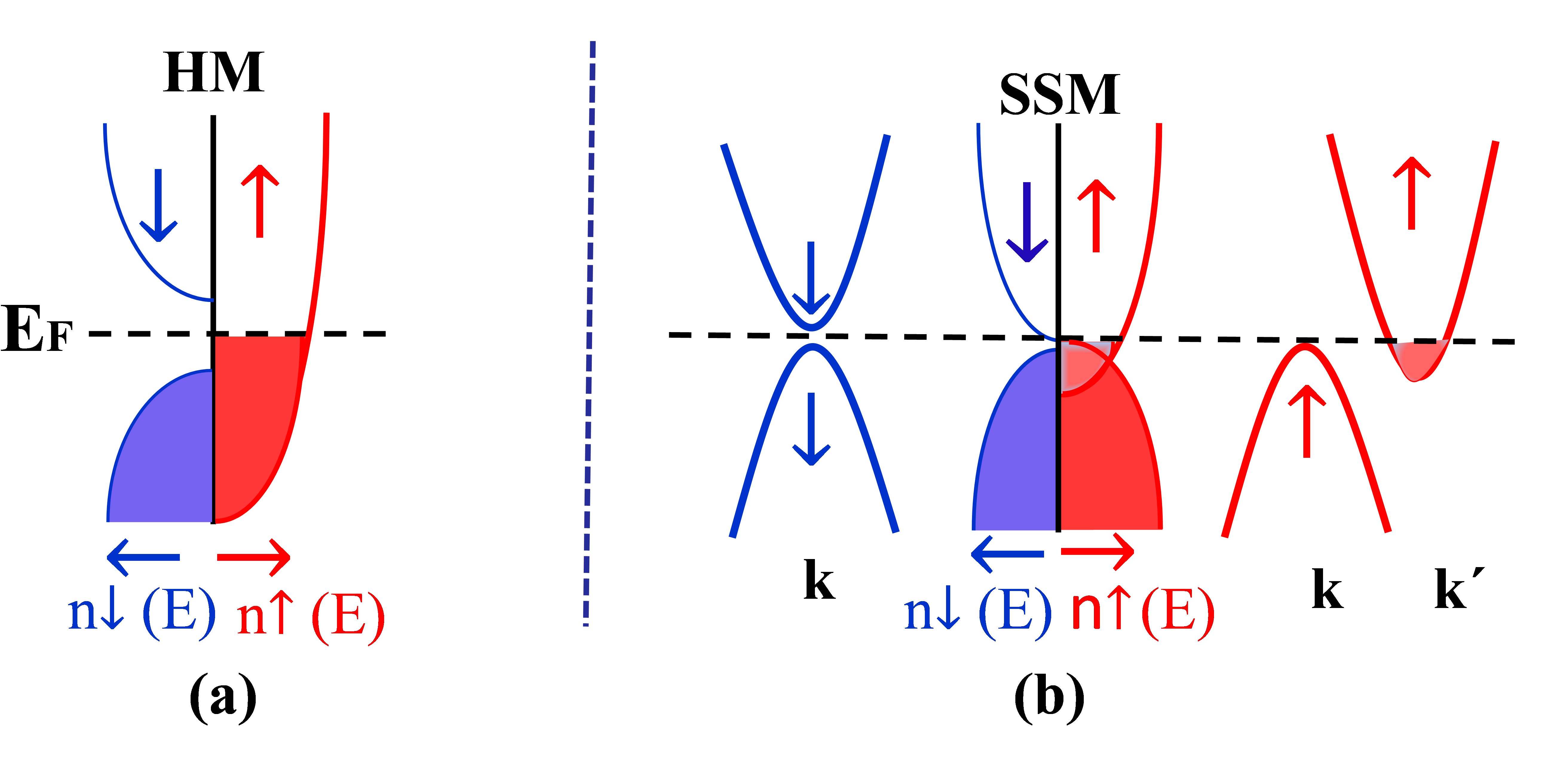}
 \caption{Schematic representation of spin polarized (a) density of states (DoS) for a conventional
half metal (HM) (b) bands (left and right panels) and DoS (middle) for spin semi-metal (SSM)}.
 \label{fig:schematic-SGS-SM}
 \end{figure}

In this article, we report the addition of a new member to the recently identified magnetic quantum material class namely spin semi-metals (SSM), with several complementary properties. This is a combined theoretical and experimental study where SSM nature is confirmed in a new quaternary Heusler alloy (QHA) CoRuVSi. The objective of this work is two-fold: (1) better understanding the key features of this relatively new class both from physics and materials perspectives, (2) highlight the importance of this class of materials for potential spintronic and thermoelectric applications.
In HMs, one of the spin bands shows metallic nature, while the other shows semiconducting/insulating behavior. SSM, on the other hand, is an unconventional class of spintronic materials in which one of the spin bands possesses semimetallic nature, while the other possesses a small gap near the Fermi level (E$_F$). Thus,  electronic states of such materials can be easily controlled  by an external perturbation (magnetic field, temperature etc.)  and hence are advantageous for spin-transport based applications. This advantage is missing in the conventional spintronic systems such as HM and SGS. A schematic representation of the density of states (DoS) and overlap of conduction and valence bands for HM and SSM are shown in Fig. \ref{fig:schematic-SGS-SM}. 

CoRuVSi is found to crystallize in the perfect cubic structure (space group $F\bar{4}3m$) with a partial L2$_1$-type disorder. The magnetization data indicates a weak ferri-/antiferro-magnetic ordering at very low temperature, with a very small saturation magnetization $\sim$ 0.13  $\mu_B$/f.u. The magnetization data indicate quenching of moment, attributed to the atomic disorder, a prediction also supported by our ab-initio disorder calculations. Theoretical studies reveal a fully compensated ferrimagnetic nature for CoRuVSi. Transport results provide strong evidence of semimetallic behavior dominated by two-band conduction, while low-T magnetoresistance data indicates the non-saturating, linear, positive magnetoresistance (LPMR), with a quadratic behavior with T. Close analysis of MR data hints toward the small-gap electronic structure near the E$_F$ as the origin of quantum LPMR, which indirectly hints toward the SSM nature present in this system. Point contact Andreev reflection (PCAR) measurements reveal a reasonably high spin polarization of $\sim$ 50\%. This matches fairly well with the theoretical calculations, again facilitating an indirect evidence of SSM feature in this system. CoRuVSi also shows a reasonably high thermopower value of $0.7$ $m Watt/ m-K^{2}$ at room temperature and hence can be further explored for its potential as a promising thermoelectric material. Our ab-initio simulation confirms the spin semimetallic feature in this alloy with a high spin polarization. Overall, the present study introduces a new member namely CoRuVSi to the magnetic quantum phase, having the potential for multifunctional applications, and gives a comprehensive analysis of the interplay between the non-trivial electronic states with magnetism and anti-site disorder. Such a combined theoretical and experimental study gives a unique platform to explore new exotic states of quantum matter.

 \section{Experimental Details}
 Polycrystalline samples of CoRuVSi were prepared using an arc melting system in a high purity Argon atmosphere using stoichiometric constituent elements having a purity of 99.99\%. To accomplish perfect homogeneity, the samples were melted several times and a very small weight loss ($<$ 0.15 \%) was observed after the final melting. To study the crystal structure, at room temperature (RT) X-ray diffraction (XRD) patterns were taken using Cu-K$\alpha$ radiation with the help of Panalytical X-pert diffractometer. For the crystal structure analysis, FullProf Suite software \cite{rodriguez1993recent} was used.                                                     
 Magnetization measurements at various temperatures were carried out using a vibrating sample magnetometer (VSM) attached to a physical property measurement system (PPMS) (Quantum Design) for fields up to 70 kOe.
Temperature and field-dependent resistivity along with the MR measurements were carried out employing a physical property measurement system (PPMS-DynaCool; Quantum Design) using the electrical transport option (ETO) in a traditional four-probe method, applying a 10 mA current at a 15 Hz frequency. Hall measurements were carried out using PPMS with the van der Pauw method by applying a 5 mA current at 21 Hz frequency. Specific heat (C$_p$) measurements were done in a 14T/2 K PPMS. A small piece of the sample (18 mg) was used to measure C$_p$, down to 2 K in zero field and in 5T applied magnetic field using a relaxation calorimetry technique.
 Thermoelectric power (TEP) in zero magnetic fields was measured using the differential dc sandwich method in a homemade setup in the temperature range of 4–300 K.
 Point contact Andreev reflection (PCAR) measurements were performed in PPMS using a superconductive Nb tip. The landing of the tip on the sample is carefully controlled by a fully automated vertical Attocube piezo-stepper. Two additional horizontal Attocube piezo steppers are used to move the sample in horizontal directions, in order to probe the pristine area of the sample. The differential conductance spectra were fitted using the modified Blonder-Tinkham-Klapwijk (m-BTK) model, as detailed elsewhere.\cite{stamenov2013point,PhysRevB.94.094415} 
 \section{Computational details}
 To study the electronic/magnetic structure of CoRuVSi, $\textit ab initio$ calculations were performed using spin-resolved density functional theory (DFT) \cite{hohenberg1964inhomogeneous} implemented within Vienna ab initio simulation package (VASP) \cite{kresse1996efficient,kresse1996efficiency,kresse1993ab} with a projected augmented-wave (PAW) basis.\cite{kresse1999ultrasoft} We used the electronic exchange-correlation potential due to Perdew, Burke, and Ernzerhof (PBE) \cite{perdew1996generalized} within the generalized gradient approximation (GGA) scheme. For the Brillouin zone integration within the tetrahedron method, a $24\times24\times24$ k-mesh was used. A plane wave energy cut-off of 420 eV was used for all the calculations. All the structures were fully relaxed with total energies (forces) converged to values less than 10$^{-6}$ eV (0.01 eV/\AA). 
The Wannier90 \cite{wannier90,PhysRevB.65.035109,RevModPhys.84.1419}simulation tool was used to compute the tight-binding Hamiltonian. A total of 62-bands were wannierized by taking projections on atomic sites as: Co (s, p, d), Ru (s, p, d), V (s, p, d), Si (s,p) etc. Further, Berry curvature, Fermi surface and anomalous Hall conductivity were calculated to investigate the semimetallic nature. The intrinsic anomalous Hall conductivity ($\sigma_{int}^{AHE}$) was estimated by integrating the Berry curvature (-$\Omega_{z} (\bf k)$) over the entire Brillouin zone considering a k-grid of $40\times40\times40$ with adoptive refinement k-mesh size of $5\times5\times5$.
To capture the effect of disorder in  L2$_1$ structure, a 64-atom special quasi-random structure (SQS)\cite{zunger1990special} was generated. SQS is a carefully generated ordered structure, which mimics the random correlations up to a certain neighboring distance in disordered compounds. To generate the SQSs, Alloy Theoretic Automated Toolkit (ATAT)\cite{van2013efficient} was used. Our generated SQSs mimic the random pair correlation functions accurately up to third-nearest neighbors.

\section{Experimental results}

\subsection{Crystal Structure}  
CoRuVSi crystallizes in LiMgPdSn prototype structure (space group $F\bar{4}3m$) with the measured lattice parameter of 5.80 {\AA} as found from the Rietveld refinement. The crystal structure can be viewed as four interpenetrating fcc sub-lattices with Wyckoff positions 4$a(0, 0, 0)$, 4$b(0.5, 0.5, 0.5)$, 4$c(0.25, 0.25, 0.25)$, and 4$d(0.75, 0.75, 0.75)$. In general, for a QHA XX$'$YZ, there exist three possible energetically non-degenerate structural configurations\cite{PhysRevB.105.144409} (keeping Z-atom at 4$a$-site) as follows:

\begin{itemize}
\item (I) X at 4$d$, X$'$ at 4$c$ and Y at 4$b$ site,
\item (II) X at 4$b$, X$'$ at 4d and Y at 4c site,
\item (III) X at 4$d$ , X$'$ at 4$b$ and Y at 4$c$ site.
\end{itemize}

\begin{figure}[t]
\centering
\includegraphics[width= 1.0\linewidth]{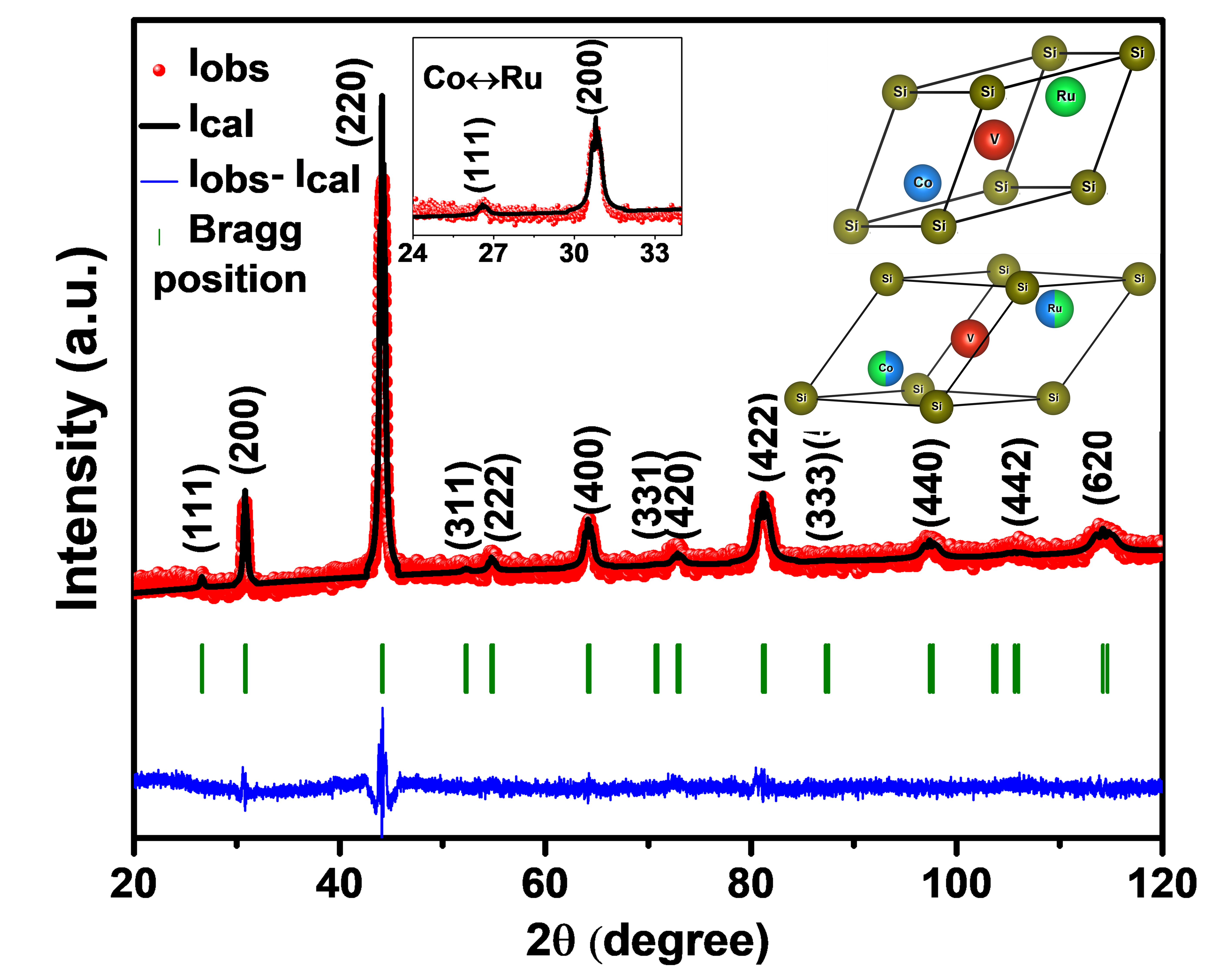}
	\caption{For CoRuVSi, room temperature powder XRD pattern including the Rietveld refined data for configuration-I with 50\% disorder between tetrahedral site Co/Ru atoms. Left inset shows a zoomed in view near super-lattice peaks (111) and (200) with L2$_1$ structure. Right insets show primitive unit cell structures corresponding to the Y-type order (top) and L2$_1$-type disorder (bottom).}
\label{fig:xrd-CFVG}
\end{figure}
 For a detailed structural analysis, we consider the structure factor for configuration-I, which can be expressed as, 
\begin{equation}
F_{hkl} = 4(f_Z + f{_Y}e^{{\pi}i(h+k+l)} + f{_X}e^{\frac{{\pi}i}{2}(h+k+l)} + f_{X'}e^{-\frac{{\pi}i}{2}(h+k+l)}).
\label{eq:sfactor}
\end{equation}
where $(h, k, l)$ are the miller indices. $f_X$, $f_{X'}$, $f_Y$, and $f_Z$ are the atomic scattering factors. The structure factor for super lattice reflections [111] and [200] can be written as:
\begin{eqnarray}
F_{111} &=& 4{[( f{_Z} - f_Y ) - i( f{_X} - f_{X'})]}\nonumber\\
F_{200} &=& 4[( f{_Z} + f_Y ) - ( f{_X} + f_{X'})]\nonumber
\label{eq:sfactor200}
\end{eqnarray}

Figure \ref{fig:xrd-CFVG} shows the room temperature XRD pattern of CoRuVSi along with the Rietveld refinement for configuration-I with 50\% disorder between tetrahedral site atoms i.e. Co/Ru (X/X$'$).  This is the best fit we got after carrying out rigorous refinement considering all possible disorders in all the configurations. Clearly, the low intensity of the superlattice peak (111) indicates the possibility of disorder in the octahedral/tetrahedral sites. For L2$_1$-type refinement, we have also  considered 50\% anti-site disorder between octahedral site atoms for configuration-I which did not fit well. The best fit with the lowest $\chi^2$ (1.80) was found in configuration-I with the L2$_1$ order (also see inset of Fig. \ref{fig:xrd-CFVG}) in comparison with other refinements considering all possible other disorders like $B2$ ($\chi^2$= 5.24), $A2$($\chi^2$ =10.5), D$O_3$ ($\chi^2$ =8.1) and perfectly ordered Y-type ($\chi^2$=4.35). As such, we conclude that CoRuVSi crystallizes in the L2$_1$ structure. The crystal structure corresponding to the Y-type order and the best fit with the L2$_1$ order are shown in the right insets of Fig. \ref{fig:xrd-CFVG}.

\subsection{Magnetic properties}
\begin{figure}[t]
\centering
\includegraphics[width= 9cm,height=6cm]{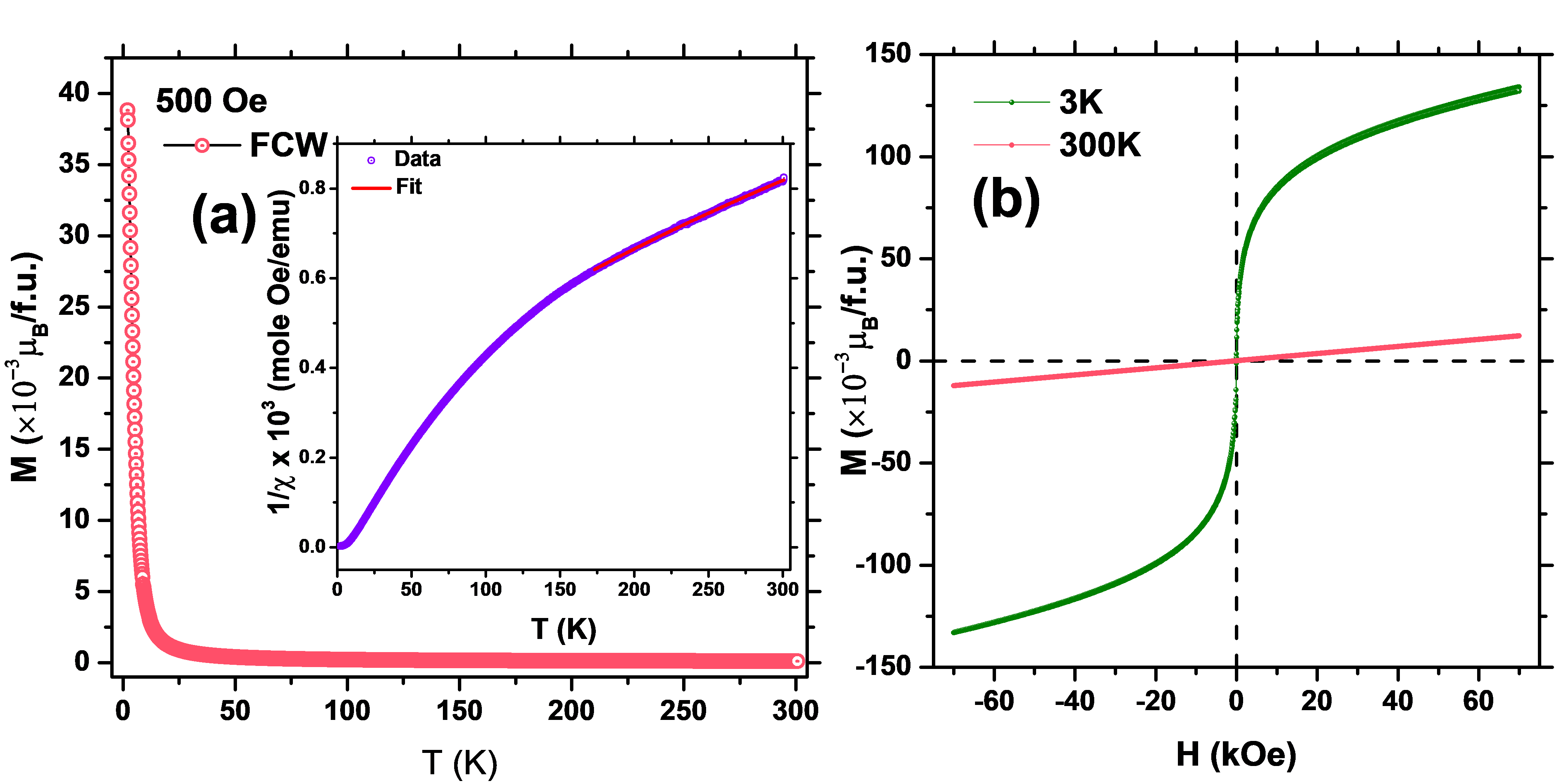}
	\caption{For CoRuVSi, (a) M vs. T in field cooled warming (FCW) mode in $H=500$ Oe. Inset shows $T$-variation of inverse susceptibility ($1/\chi$) along with Curie-Weiss fitting in high $T$-regime. (b) M-H curves at 3K and 300K.}
\label{fig:mt-CFVG}
\end{figure}

Figure \ref{fig:mt-CFVG}(a) shows magnetization (M) vs. temperature (T) for CoRuVSi measured at H $=500 $ Oe. The field cooled warming (FCW) curve taken at H$=500 $Oe shows a rapid increase in M below 25 K, which hints toward the possibility of magnetic ordering at very low T.
Inset of Fig. \ref{fig:mt-CFVG}(a) shows the Curie-Weiss (C-W) law fitting of the susceptibility data in high T range at H=500 Oe.

The magnetic moment ($m$) can be calculated considering the Slater-Pauling (S-P) rule using the total number of valence electrons ($n_v$) of the constituent elements.\citep{graf2011simple} The total moment ($m$) per formula unit can be expressed as:\cite{ozdougan2013slater,zheng2012band} $ m = (n_v - 24) \ \ \ \mu_B/f.u. $
For CoRuVSi, the S-P rule predicts $ m$ =2.0 $\mu_B$/f.u. in the fully ordered state, but interestingly, the M-H curve shows a very small saturation magnetization ($0.13$ $\mu_B$/f.u.) even at 3 K, which is a complete deviation from the S-P rule. The presence of L2$_1$ disorder can be a plausible reason for the quenching of moment in this system. To get an idea about the magnetic interactions present in this system, the inverse-susceptibility data (H=500 Oe) has been fitted (solid red line) above 150 K using the C-W law ($\chi^{-1}=\frac{1}{\chi_0 + C/(T-\theta_P)}$) and from the fitting, we obtained effective moment $ m$ =0.2 $\mu_B$/f.u., $\chi_0$=$0.40$ $emu/mole-$Oe and Weiss temperature, $\theta_P$= $-$93 K, that indicates the presence of antiferromagnetic interactions in the system. The sharp shoot below 25 K may arise because of the moments of ferri/antiferromagnetic clusters can easily prevail over the paramagnetic regime at low T.
Additionally, non-saturating behavior (up to 70 kOe field) of low-T M-H curve (Fig. \ref{fig:mt-CFVG}(b)), along with no hysteresis indicates superparamagnetic-like behavior in this system, attributable to the L2$_1$ disorder, which gives rise to the moment quenching.
Thus, magnetization data reveal the possibility of small magnetic clusters, formed by weakly interacting moments, with no spontaneous magnetization. This confirms the absence of coherent long-range ordering, mediated by the atomic disorder, giving rise to complex magnetic nature in CoRuVSi.\cite{PhysRevLett.59.586}


\subsection{Transport properties}
\subsubsection{PCAR}
\begin{figure}[t]
\centering
\includegraphics[width= 1.0\linewidth]{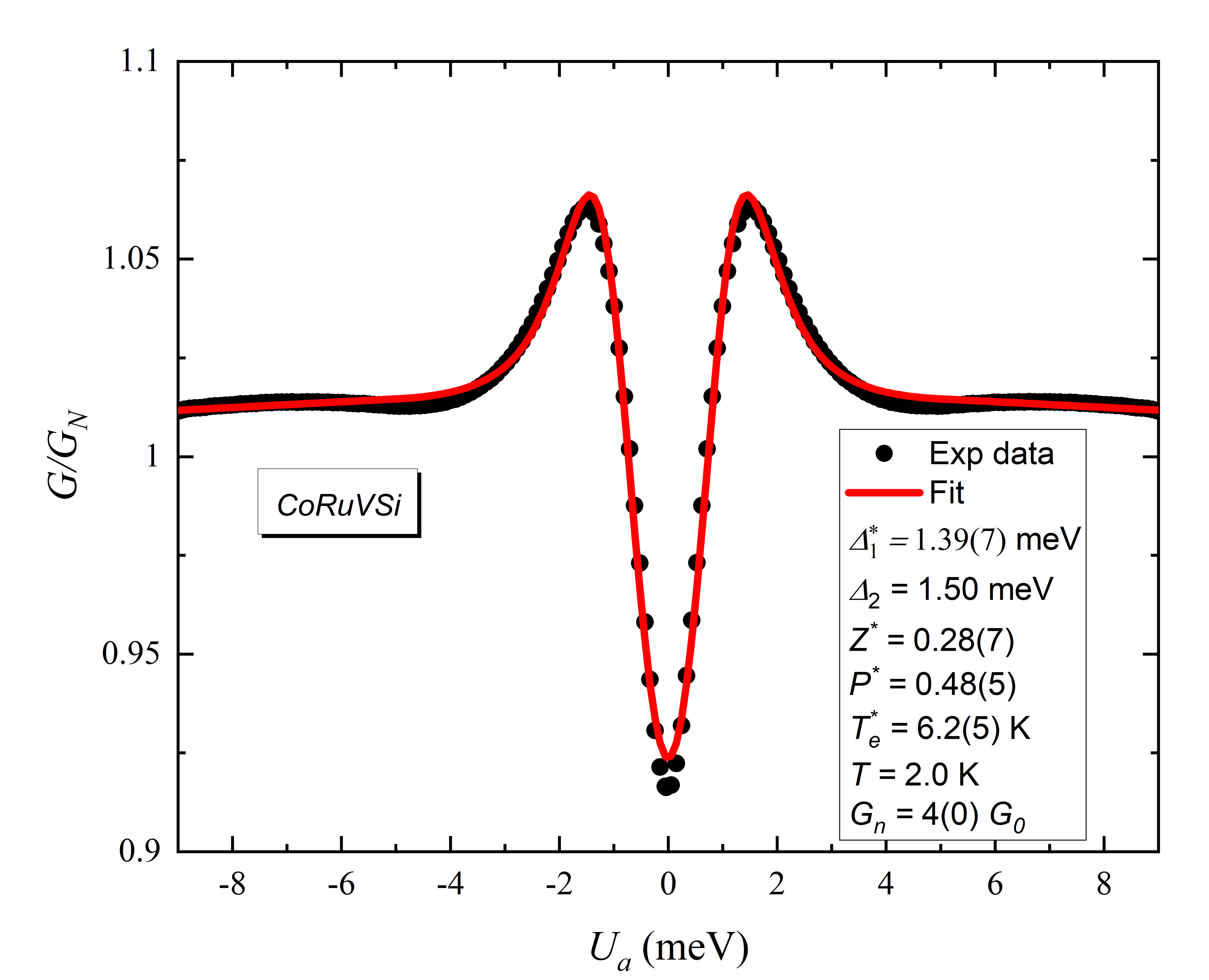}
	\caption{For CoRuVSi, point-contact Andreev reflection (PCAR) spectra, along with fit, and extracted parameters. The extracted m-BTK-model parameters are provided inside the box. The measured spin polarization was found to be $P$= 48(5)\%.}
\label{fig:ac}
\end{figure}

The electronic spin polarization $P$ at the Fermi level ($E_F$) is defined as:
	
	\begin{equation}
	P = \frac{{n_{\uparrow}(E_F)} - {n_{\downarrow}(E_F)}}{{n_{\uparrow}(E_F)} + {n_{\downarrow}(E_F)}}
	\end{equation}
where ${n_{\uparrow}(E_F)}$ and ${n_{\downarrow}(E_F)}$ are the spin-projected density of states at ($E_F$) for spin-up and -down channels respectively. Figure \ref{fig:ac} summarize the spin polarization data as obtained by the PCAR measurement. It shows a maximum spin polarization of $\sim$ 50\% at the Fermi level, which is reasonably high to serve as potential spintronic material.\cite{bainsla2015spin} \textcolor{black}{The reduction in spin polarization value as compared to the theoretical value corresponds to a narrowing of the spin gap in the density of states, which is possibly due to the presence of small density of states attributed to the disorder in the real system.}

\subsubsection{Resistivity}
\begin{figure}[t]
\centering
\includegraphics[width=1.0\linewidth]{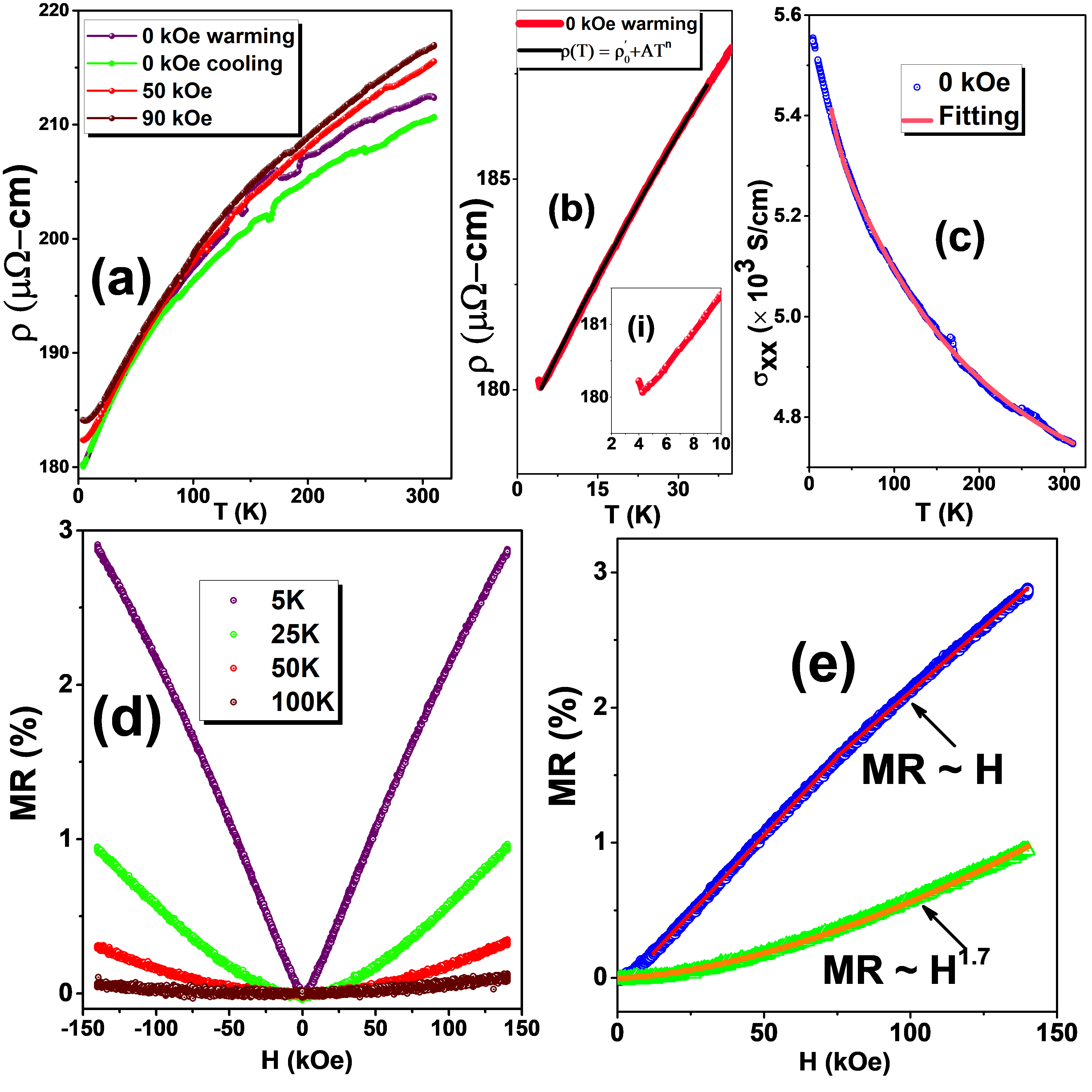}
	\caption{ For CoRuVSi, (a) Longitudinal resistivity ($\rho_{xx}$) vs. T in three different fields, 0, 50 and 90 kOe. (b) $\rho(T)=\rho_0 + AT^{n}$ fitting in the T-range 5-30 K. Inset shows a zoomed in view of the  $\rho_{xx}$ data in the low-$T$ range. (c) longitudinal conductivity ($\sigma_{xx}$) vs. $T$ in zero field along with a two-carrier model fit between 30-310 K. (d) MR vs. $H$ at four different temperatures 5, 25, 50,  and 100K. (e) linear and quadratic fitting of MR vs. H at 5 K and 25 K respectively.}
\label{fig:RT-CFVG}
\end{figure}

Figure \ref{fig:RT-CFVG}(a) shows the T-dependence of resistivity ($\rho_{xx}$) at different applied fields. It reflects a semi-metallic behavior (also revealed by the electronic structure calculations shown later). To gain further understanding, we have fitted the zero-field resistivity data considering various scattering mechanisms in various T-ranges. A dip-like feature below 5 K in the zero-field data indicates the possibility of weak localization arising from disorder in this system. Resistivity follows a power law behaviour ($\rho(T)=\rho_0 + AT^{n}$) in the T-range of 5 K$<$T$<$30 K, as shown in Fig. \ref{fig:RT-CFVG} (b). Above 30 K, resistivity data fit well with the two-carrier model which supports the semi-metallic behavior in this system (also supported by the carrier concentration from Hall data, shown later). 
For further investigation, we have fitted the conductivity ($\sigma$) data (see Fig. \ref{fig:RT-CFVG}(c)) with a modified two-carrier model (Eq. \ref{eq:tbm-final}),\cite{kittel2007introduction,jamer2017compensated} in the T-range $30-310$ K.  
A two-carrier model for $\sigma$ can be written as,
\begin{equation}
\sigma(T) = e (n_e \mu_e + n_h \mu_h)
\label{eq:tbm}
\end{equation}
where, $n_i=n_{i0}\ e^{-\Delta E_i/k_\mathrm{B}T}$($i=e,h$) are the electron/hole carrier concentrations with mobilities $\mu_i$ and pseudo-energy gaps $\Delta E_i$. 
Eq.(\ref{eq:tbm}) can be further expressed as,
\begin{equation}
\sigma(T) = [A_e(T) \ e^{-\Delta E_e/k_\mathrm{B}T} + A_h(T) \ e^{-\Delta E_h/k_\mathrm{B}T}].
\label{eq:tbm-final}
\end{equation}
After fitting the conductivity data with the above equation, we obtained the pseudo-energy gaps for electrons and holes to be 0.11 meV and 15.9 meV, which are quite small and resemble those of a narrow band gap semiconductor. It appears that the atomic disorder plays a crucial role in significantly reducing the pseudo-gaps, especially for electrons which have an extremely small gap and are likely to become metallic with small perturbations (e.g. applied field, thermal fluctuation etc.).

\subsubsection{Magnetoresistance}
Figure \ref{fig:RT-CFVG}(d) shows the field dependence of MR at different T, where MR is defined as MR(H)=$ \left[ \rho(H) - \rho(0)\right]/\rho(0)$ $\times 100\%$. At 5 K, non-saturating linear positive magnetoresistance (LPMR) is observed, which is also confirmed by the linear fitting of MR vs. H (see Fig. \ref{fig:RT-CFVG} (e)). But with increasing T, field-dependent MR(H) becomes almost quadratic in nature and at 25K an unsaturated quadratic MR is observed. The origin of LPMR at the lowest T (5 K) is possibly due to the zero/small-gap electronic structure near the E$_F$.\cite{PhysRevB.103.104427} MR magnitude decreases gradually with T and at 100 K it becomes almost zero. To characterize the type of carriers, we have further performed the Hall measurement (as described below).

\subsubsection{Hall Measurements} 
\begin{figure}[t]
\centering
\includegraphics[width=1.0\linewidth]{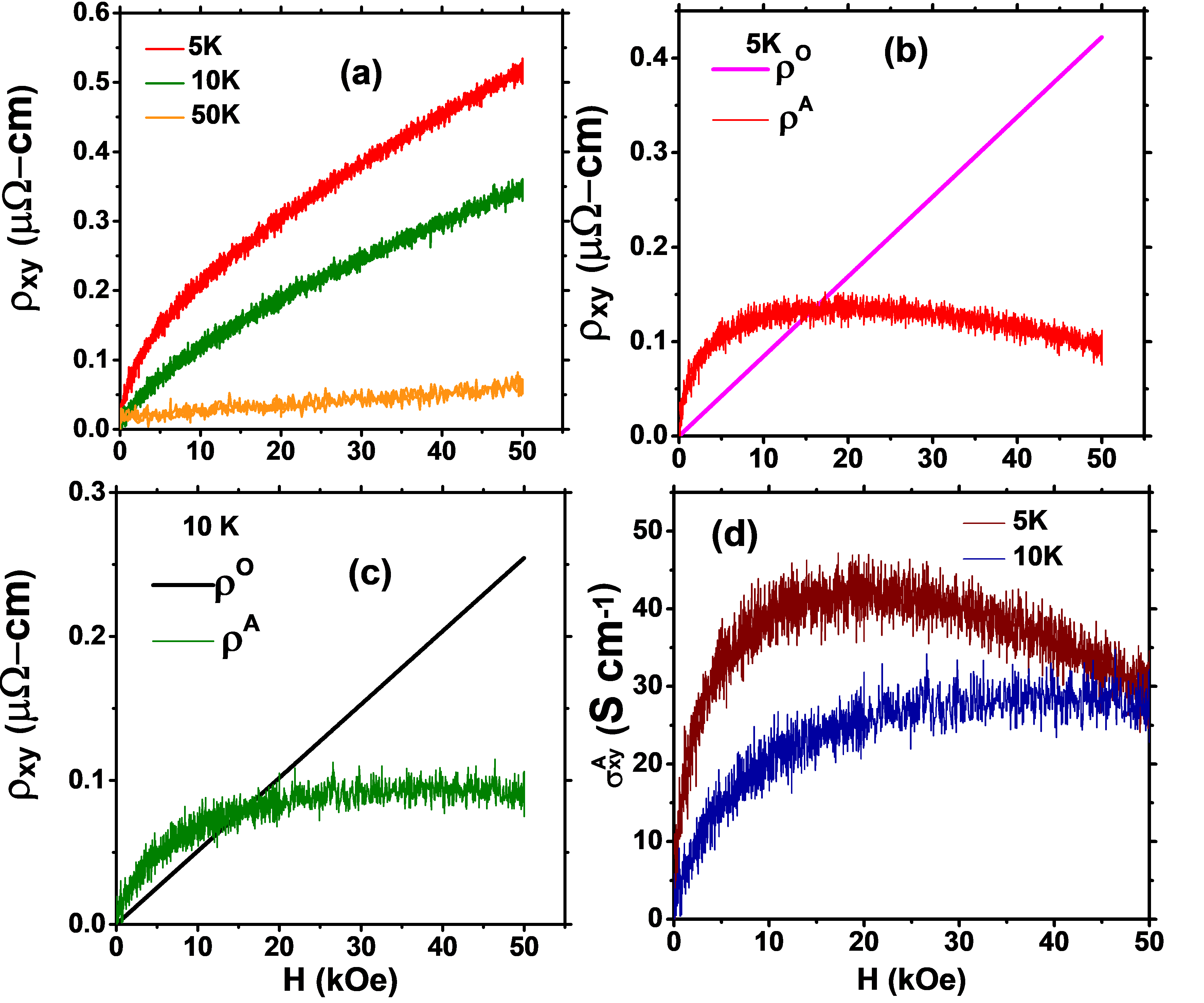}
\caption{For CoRuVSi, (a) Hall resistivity ($\rho_{xy}$) vs. applied field (H) at 5, 10 and 50 K. (b-c) Ordinary and anomalous contributions of $\rho_{xy}$ at 5 and 10 K respectively. (d) Anomalous Hall conductivity ($\sigma_{xy}^{A}$) vs. H at 5 and 10K.}
\label{fig:Hall-CFVG}
\end{figure}

Figure \ref{fig:Hall-CFVG}(a) shows the field-dependence of Hall resistivity $\rho_{xy}$ at various T. Generally, Hall resistivity for a magnetic material has two contributions expressed as,
\begin{equation}
\rho_{xy}(T)=\rho_{xy}^{O} + \rho_{xy}^{A}=R_{0}H+R_{A}M,
\label{eq:Hall}
\end{equation}
where, $\rho_{xy}^{O}$ and $\rho_{xy}^{A}$ are ordinary and anomalous contribution to $\rho_{xy}$, $R_0$ and $R_A$ denote the ordinary and anomalous Hall coefficients respectively. At 5 K and 10 K, both the contributions are observed, but at 50 K, the amplitude of $ \rho_{xy}$ drops abruptly to zero (see Figs. \ref{fig:Hall-CFVG}(b-c)), as anomalous contribution die out with increasing T and only ordinary contribution remains. We have extracted $\rho_{xy}^{O}$ and $\rho_{xy}^{A}$ contributions at 5 K and 10 K, as shown in Fig. \ref{fig:Hall-CFVG}(b-c).
To scale the anomalous Hall effect (AHE) contribution, we have fitted a linear curve to $\rho_{xy}$ data at large H, and extracted $\rho_{xy}^{A}$ contribution. From the AHE data, a small AHE contribution ($\sim$ $0.15 \mu\Omega$-cm) is observed.  Figure \ref{fig:Hall-CFVG}(d)  shows the field-dependence of anomalous Hall conductivity ($|\sigma_{xy}^A|$ $\sim$ $\frac{\rho_{xy}^A}{\rho_{xx}^2}$) at 5 K and 10 K. $\sigma_{xy}^{A}$ reaches a maxima $\sigma_{xy0}^{A}=$45 S cm$^{-1}$ at 5 K, confirming a non-saturating and non-linear behaviour. The measured value of the carrier concentration$(n)$ at 5 K is 7.4$\times10^{18}$ cm$^{-3}$, which falls well within the range of carrier densities for semimetals/semiconductors, again indicating the semi-metallic nature of CoRuVSi.\cite{chen2021large} The positive slope of $R_0$ reveals holes as majority charge carriers. The origin of this behavior may be attributed to the change in electronic structure brought about the atomic disorder.

\subsubsection{Thermoelectric power }
\begin{figure}[t]
	\centering
	\includegraphics[width=1.0\linewidth]{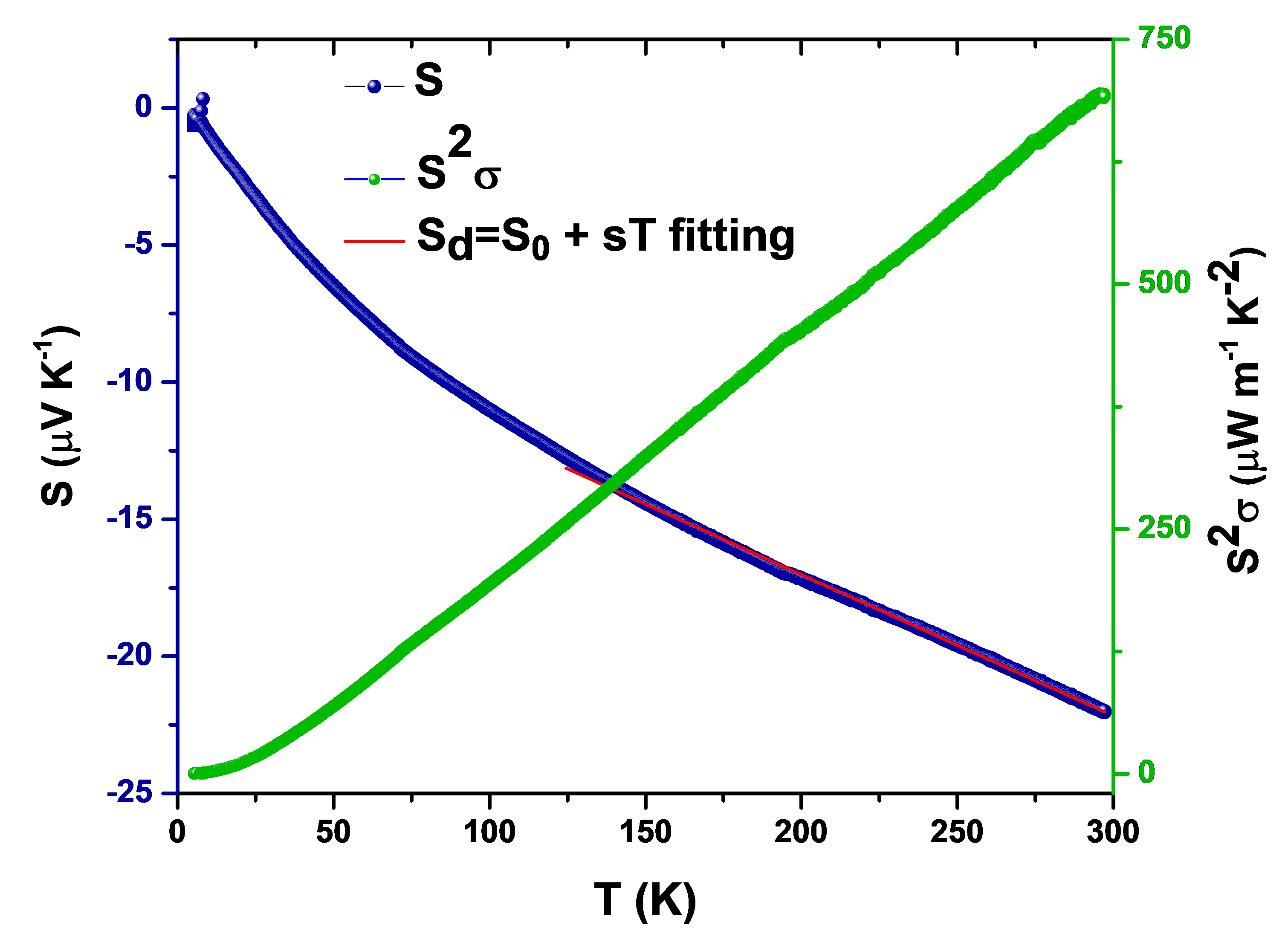}
	 \caption{For CoRuVSi, $T$-dependence of thermoelectric power (S) and power factor ($S^2\sigma$) along with the fitting of diffusion thermopower (S$_d$=S$_0$+s$T$) between 100-300 K.}
	\label{fig:TEP}
\end{figure}

Figure \ref{fig:TEP} shows the T-dependence of the Seebeck coefficient (S) (left y-scale) along with the power factor ($S^2\sigma$) (right y-scale). S shows a sub-linear variation with T, which is typically seen in semimetals \cite{pan2021thermoelectric,sk2022experimental} The negative slope of S with T corresponds to electron-driven thermopower, which again reveals two-carrier conduction in this system. The linear behavior of S suggests the dominance of diffusion thermopower. $\vert$S$\vert$ attains a value of 23 $\mu V/K$ at 300 K, which is comparable to that of other potential thermoelectric (TE) materials, at RT.\cite{yu2009high,hayashi2017structural,lue2002thermoelectric}
To further evaluate the potential of CoRuVSi for thermoelectric  applications, we have calculated the power factor (PF=S${^2}\sigma$), a key parameter determining the efficiency of thermoelectric material. PF varies linearly with T and attains a maximum value of $0.7$ $m Watt/m-K^{2}$ at RT, which is reasonably high as compared to many Heusler-based TE materials,\cite{yan2011enhanced} and also comparable with other reported promising TE materials.\cite{hinterleitner2019thermoelectric,huang2015new,fu2013electron,PhysRevB.103.085202,PhysRevB.105.144409} To get an idea about the carrier density (n) and E$_F$, S-data is fitted with the equation S$_d$=S$_0$+s$T$ in the high-T regime, where S$_d$ is the diffusion thermopower, S$_0$ is a constant and s$=\frac{\pi^2k_{B}^{2}}{3e\text{E}_F}$. From this fitting, we obtained E$_F$ = $1.41$ eV and n=$7.2\times{10^{18}}$cm$^{-3}$, which is comparable with the Hall data and falls well within the range of carrier densities of promising TE materials. 
Further investigation on high T measurements and thermal transport can help in determining the potential of CoRuVSi as a promising thermoelectric material at high T.

\subsubsection{Specific heat}
\begin{figure}[b]
\centering
\includegraphics[width=1.0\linewidth]{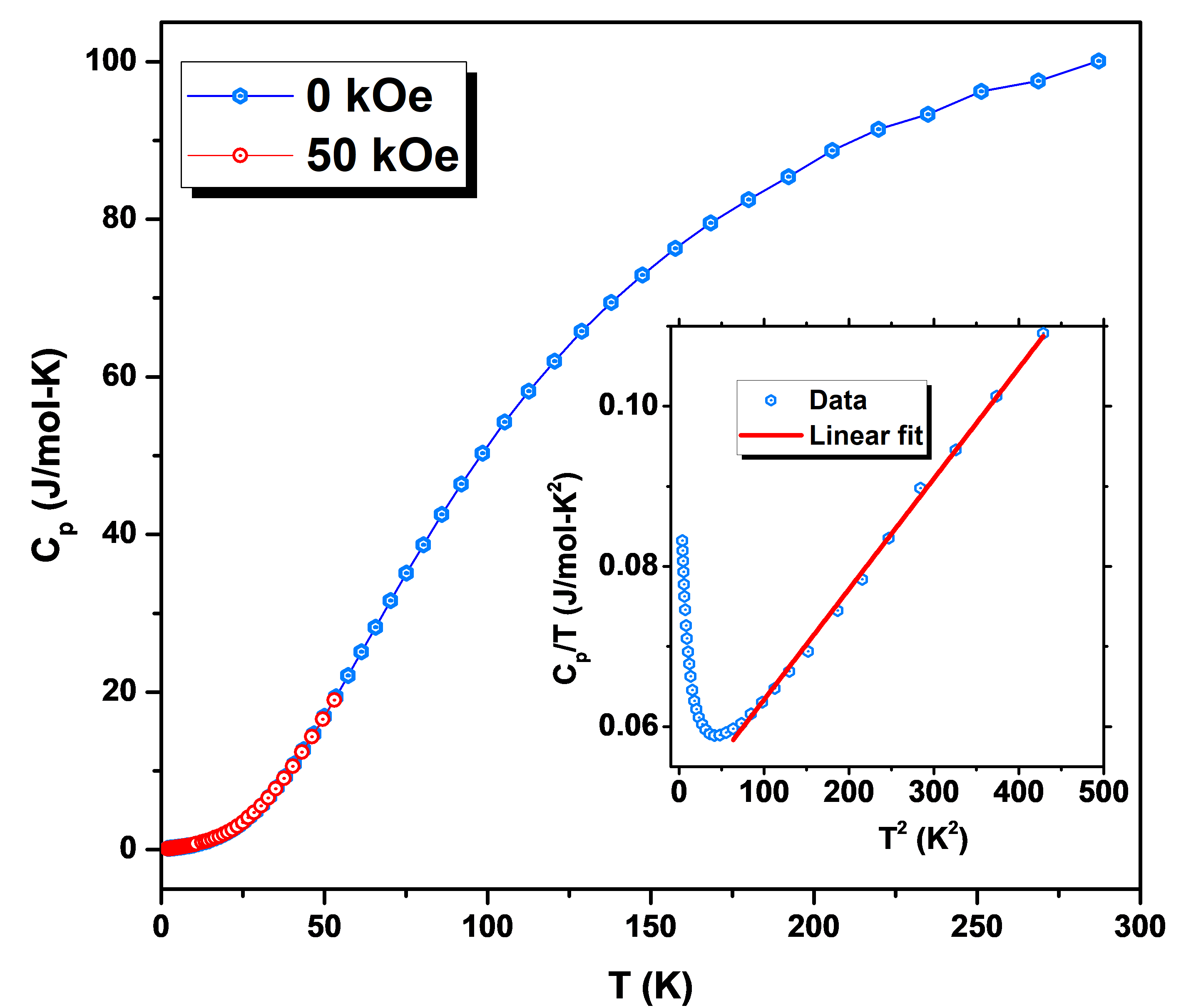}
	\caption{For CoRuVSi, specific heat (C$_p$) vs. $T$ for 0 and 50 kOe fields. Inset shows {C$_p$/T} vs. $T^2$ along with the linear fit (solid red line) for zero field.}
\label{fig:hc}
\end{figure}

Figure \ref{fig:hc} shows the T-dependence of specific heat for 0 and 50 kOe. The low-T C$_p$ data is fitted with the equation $C(T)=$ $\gamma$$T + $$\beta$T$^3$, where the first term is electronic contribution to C$_p$ while the second term is the low-T phonon contribution. The inset of  Fig. \ref{fig:hc} shows {C$_p$/T} vs. T$^2$ plot along with the linear fit. From this fitting, we obtained $\gamma$=0.05 $J/mole-K^{2}$ (Sommerfeld coefficient), which in turn gives the density of states at E$_F$ i.e. $n(E_F)$=$3\gamma/(\pi^{2}k_{B}^{2})$ $\sim$ 4.5 states/eV f.u.\cite{venkateswara2019coexistence} This value matches quite well with the theoretical results (see next section) and is in good agreement with small DoS near E$_F$ for semimetals. This is another indication of expected semimetallic feature in CoRuVSi. From the fitting, we also extracted the Debye temperature, $\theta_D$= 383 K using the value of $\beta$=1.3795$\times$ $10^{-4}$ $J/mole-K^{4}$. Interestingly {C$_p$/T} vs. T$^2$ plot shows a shallow minimum, which may be related to the AFM-like interaction present in CoRuVSi.

\section{Theoretical Results}
\begin{figure}[t]
\centering
\includegraphics[width= 1.0\linewidth]{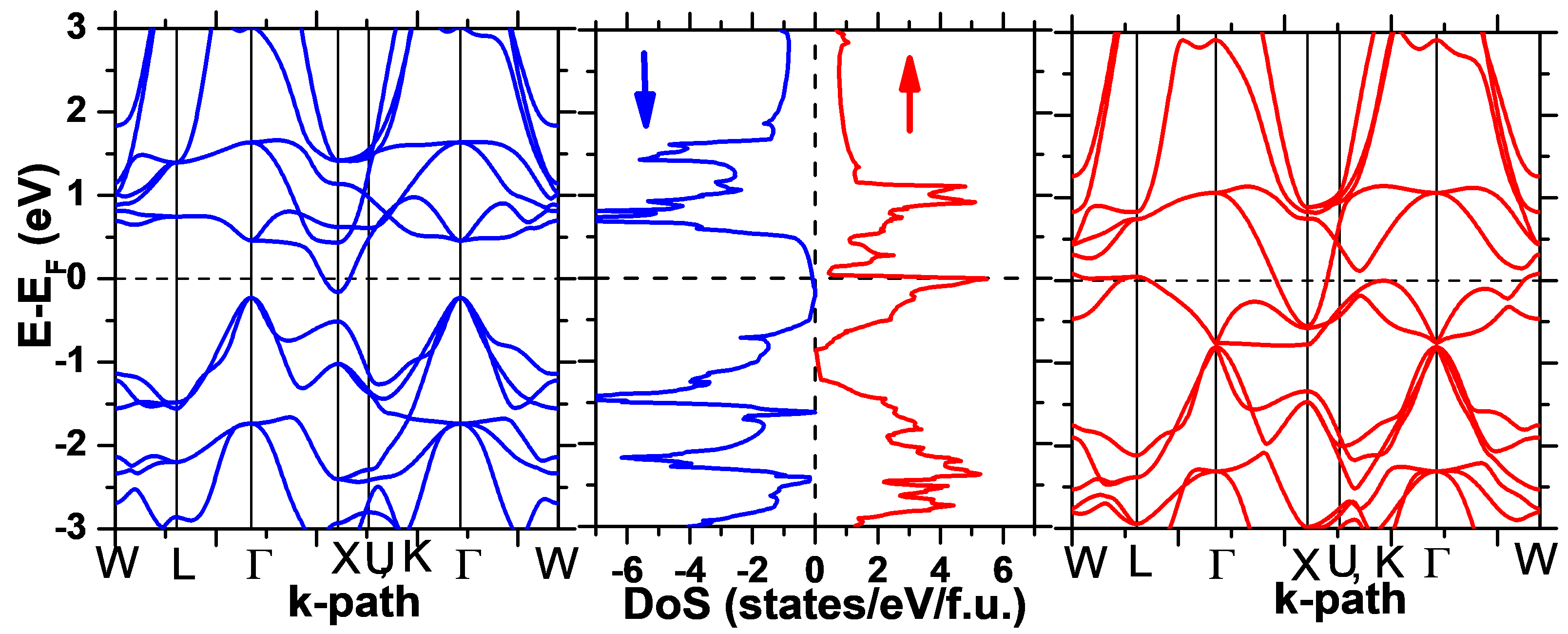}
\caption{For ordered CoRuVSi (in type-I configuration), spin resolved band structure and density of states (DoS) at the optimized lattice parameter ($a_0$). A few electron pockets at/around the E$_F$ are observed for minority spin-channel.}
\label{fig:CRVS-band}
\end{figure}

\begin{table}[b]
\centering
\caption{For ordered CoRuVSi, theoretically optimized lattice parameter ($a_0$), total and atom-projected magnetic moments ($ \mu_B$), and relative energy ($\Delta E$)  for type I, II and III configurations with respect to the energy of type-I configuration.}
\begin{tabular}{l c c c c c c}
\hline \hline
Type&  $a_0$ (\AA) $ \ $  &  $m^{\mathrm{Co}}$ & $\ $ $m^{\mathrm{Ru}}$ $\ $  &  $m^{\mathrm{V}}$  & $\ $ $m^{\mathrm{Total}}$ $ \ $ & $\Delta E$(eV/f.u.) \\ \hline 
I   &  5.82   & 1.68 	& 	 -0.13  	& 	 0.53 	& 2.07	&  0   \\
II  &  5.85  & 1.32		& 0.73	& 	-0.21 	&  1.84		& 0.32    \\
III    & 5.83  &  1.35  	&     0.28 	&	-0.5 	& 1.16		&  0.22  \\
\hline \hline
\end{tabular}
\label{tab:theory-CFVG}
\end{table}
We have used {\it ab-initio} simulation to investigate various magnetic states including para-, ferro-, antiferro-, and ferri-magnetic configurations in the ordered and L2$_1$-disordered phases for CoRuVSi. Out of all the configurations, type-I configuration (see Sec. IV(A)) with ferrimagnetic ordering turned out to be energetically the most favorable one. Table \ref{tab:theory-CFVG} shows the optimized lattice parameters, total and atom-projected moments and relative energies of three different ordered structures (type-I, II and III) in their respective lowest energy magnetic ground state. Figure  \ref{fig:CRVS-band} shows the spin-resolved density of states and band structure for the lowest energy type-I configuration, which indicates a nearly half-metallic ground state with a net magnetization of $\simeq$2\ $\mu_B/f.u. $.

\begin{figure}[t]
\centering
\includegraphics[width=\linewidth]{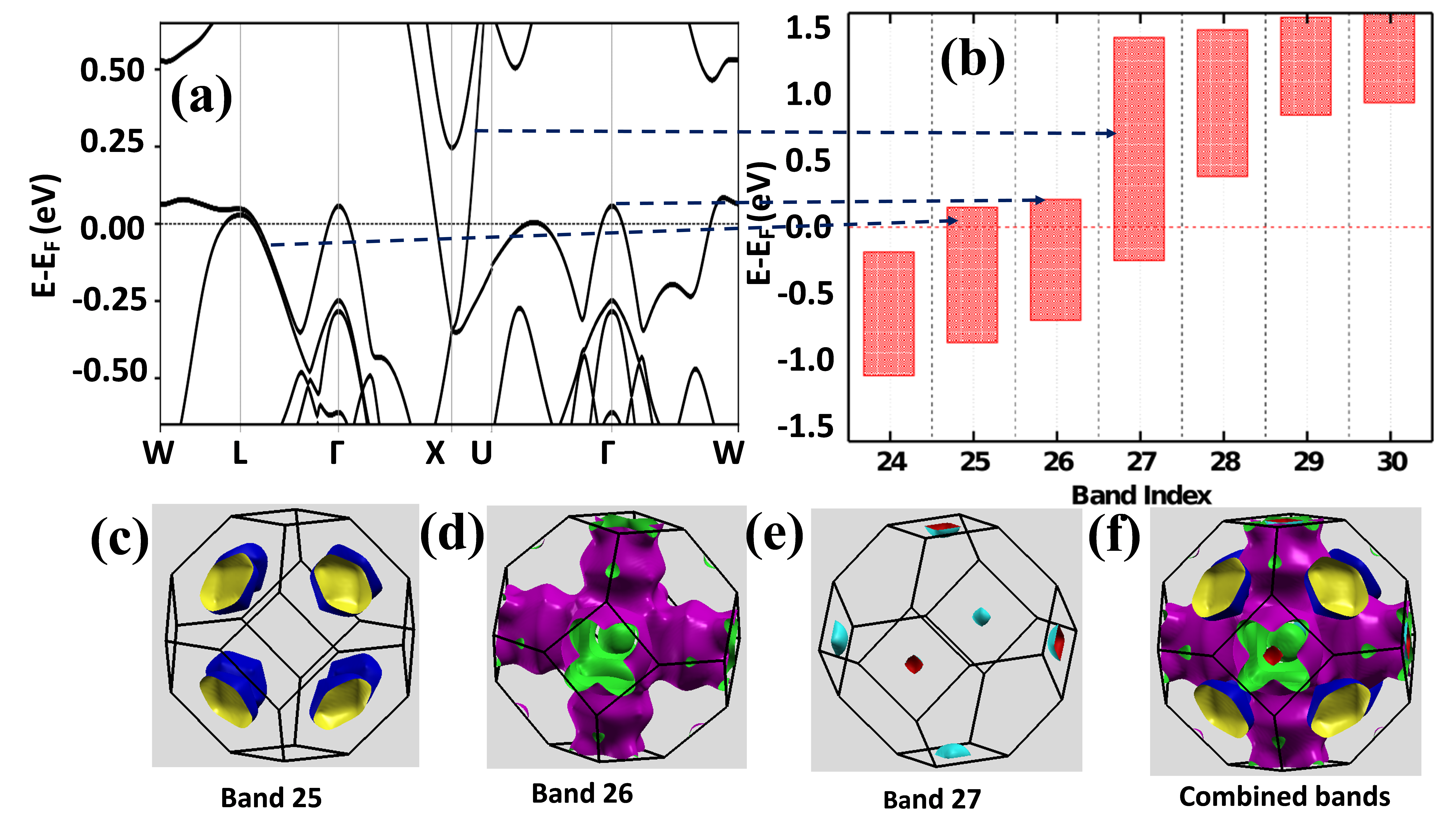}
\caption{For ordered CoRuVSi (in type-I configuration), (a) band structure including spin-orbit coupling, (b) widths of various bands, illustrating the band-crossing near $E_F$. Fermi surfaces attributed to (c) band 25, (d) band 26, (e) band 27, and (f) combined bands, illustrate the emerging electron and hole pockets at the high symmetry points confirming the semimetallic feature.}
\label{fig:w90-1}
\end{figure}

\begin{figure}[b]
\centering
\includegraphics[width=0.85\linewidth]{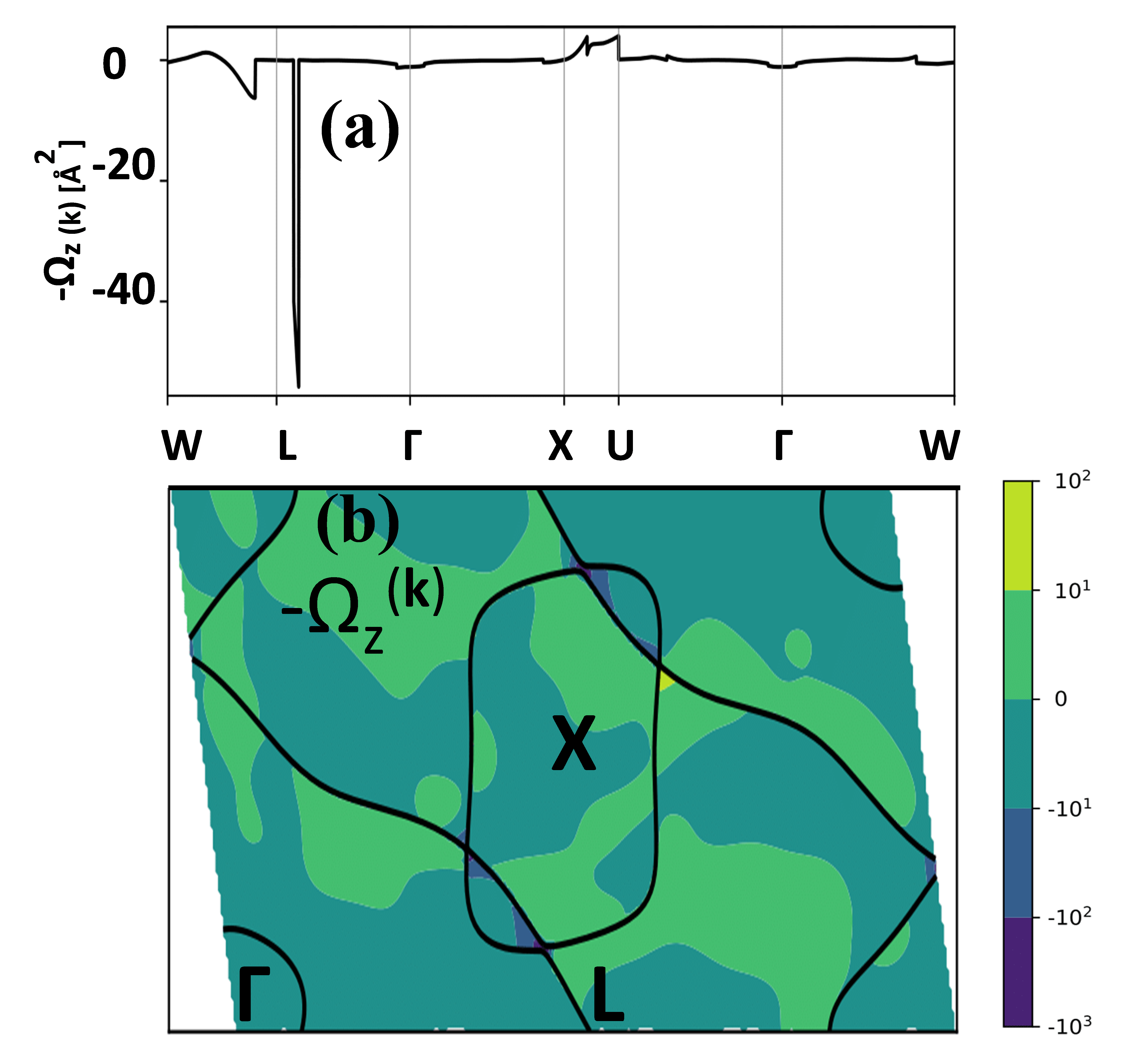}
\caption{For ordered CoRuVSi (in type-I configuration), simulated Berry curvature (-$\Omega_{z}(\bf k)$) (a) along the high symmetry paths and (b) in the k$_x$-k$_y$ plane at E$_F$. Black solid lines show intersections of the Fermi surface with this plane. }
\label{fig:w90-2}
\end{figure}

In order to further explore the half metallic/semimetallic nature, we have simulated the band structure of the ordered CoRuVSi (type-I configuration) including the spin-orbit coupling (SOC), as shown in Fig. \ref{fig:w90-1}(a). The corresponding band crossing near the Fermi level are shown in Fig. \ref{fig:w90-1}(b). This clearly illustrates the semimetallic nature with the bands 25, 26 and 27 crossing E$_F$. The Fermi surfaces corresponding to these three individual bands as well as the net combined Fermi surface are shown in Fig. \ref{fig:w90-1}(c-f).  This confirms the emergence of hole pockets from bands 25 and 26, while band 27 gives rise to electron pocket. Figure \ref{fig:w90-2}(a) shows the z-component of the Berry curvature ($\Omega_z(\mathbf{k})$) along the high symmetry $\vec{k}$-points. The corresponding 2D projection of $\Omega_z(\mathbf{k})$ in k$_x$-k$_y$ plane is shown in Fig. \ref{fig:w90-2}(b). Here, black solid lines show intersections of the Fermi surface with this plane.
The large spike in the Berry curvature near the vicinity of $L$ point is attributed to the two spin-semimetallic bands (25 and 26), one of which is unoccupied (band 25) and the other (band 26) is occupied in a small k-interval. Due to spin-orbit coupling, a small energy gap opens up, giving rise to a small energy denominator in the definition of  Berry curvature (i.e. $\Omega_n$ $\sim$ $1/(\Delta {\varepsilon_n}^2)$ from the Kubo-formula)\cite{PhysRevLett.92.037204}. So, these topological spin-semimetallic bands induce an appreciable Berry curvature, which is purely intrinsic in nature. The intrinsic anomalous Hall conductivity is calculated by integrating $\Omega_z(\mathbf{k})$ over the entire Brillouin zone (BZ), using the following expression,\cite{PhysRevLett.92.037204}
\begin{equation}
\sigma_{int}^{AHE}= - (e^2)/(8\pi^3\hbar)  \int_{(BZ)} d^3k\ \Omega_z(\mathbf{k}),
\end{equation}

\begin{figure}[h!]
\centering
\includegraphics[width= 1.0\linewidth]{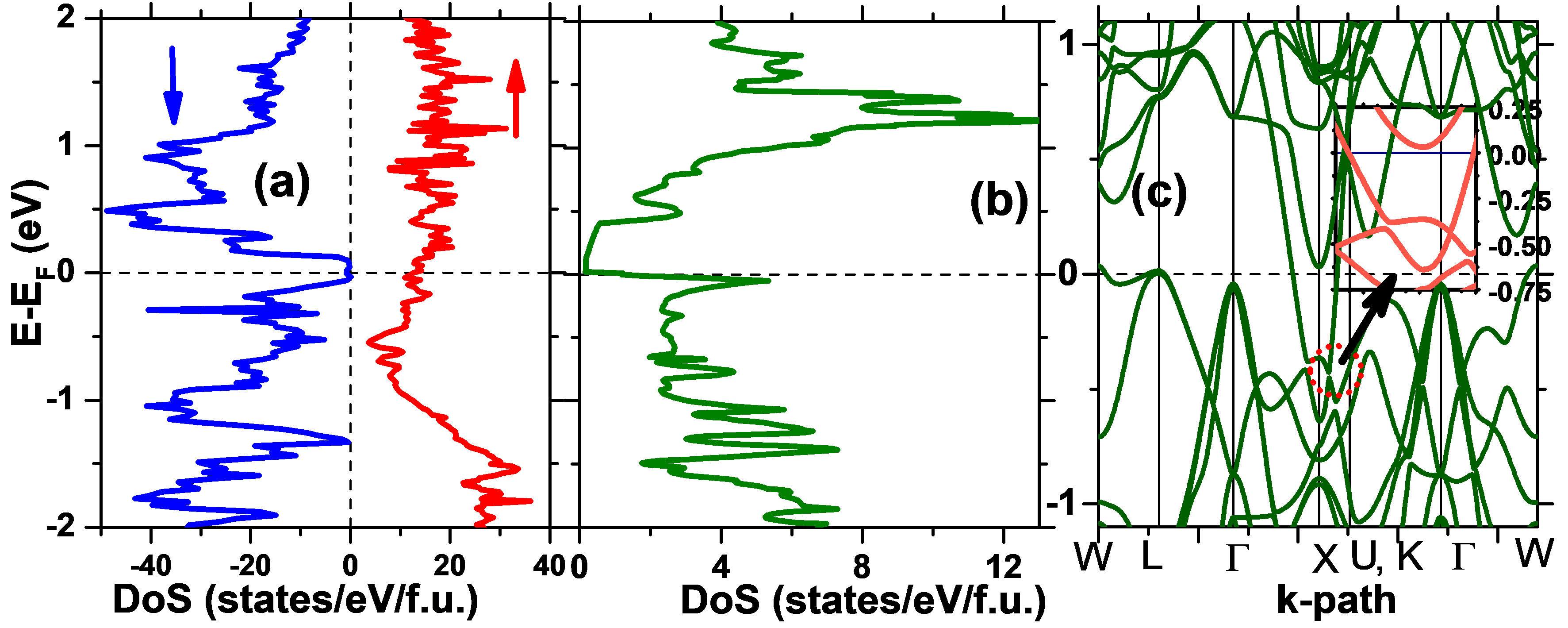}
\caption{For L$2_1$ partially ordered CoRuVSi (SQS structure) (a) spin resolved density of states without spin orbit coupling (SOC) at the optimized lattice parameter($a_0$) (b-c) DoS and band structure including the effect of SOC. Red circle in Fig. (c) highlights the linear band crossing at/near $X$ point with the band inversion, again confirming semimetallic nature. Inset shows a zoomed in view of this band crossing.}
\label{fig:soc}
\end{figure}

\begin{figure}[b]
\centering
\includegraphics[width=\linewidth]{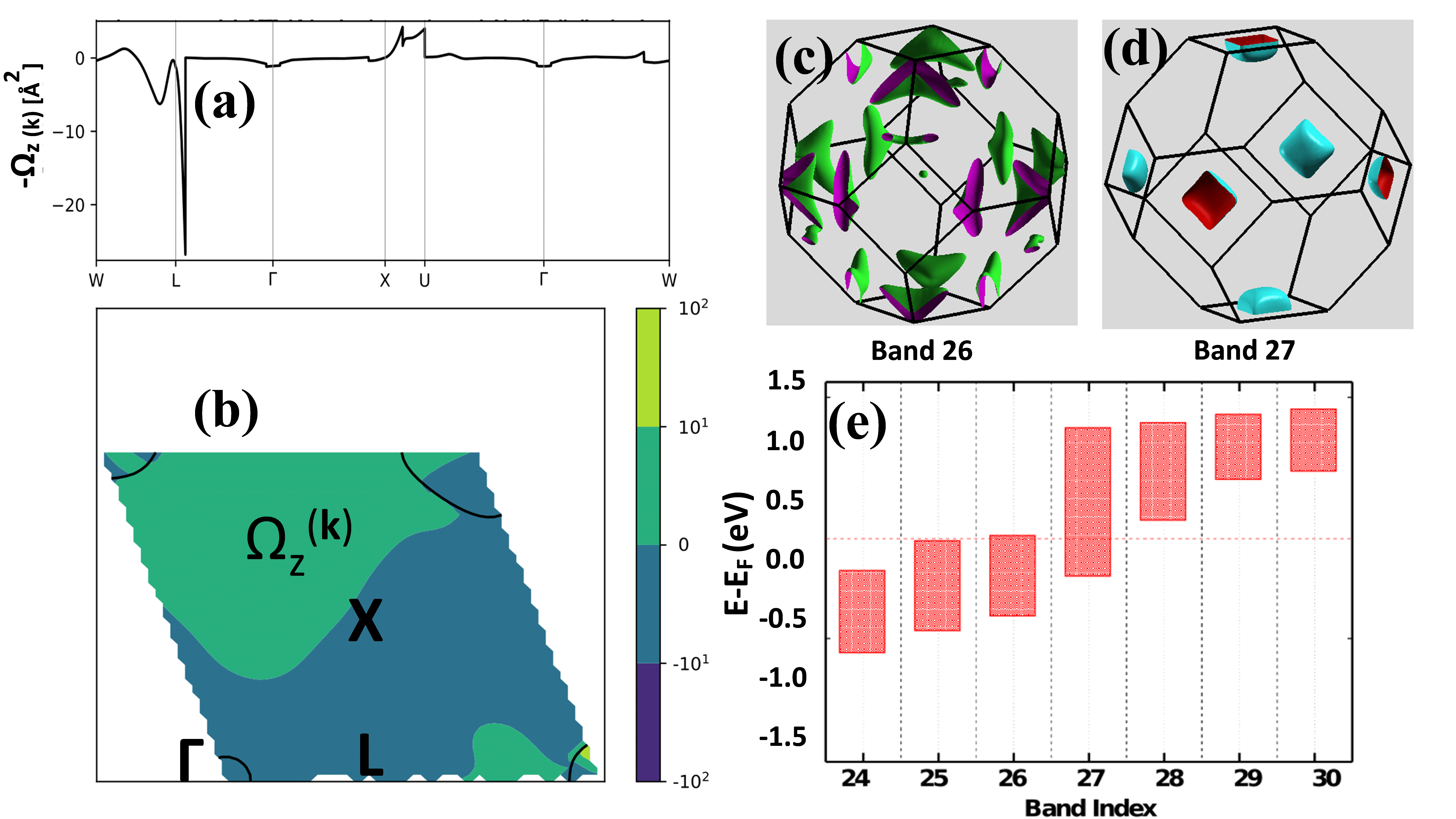}
\caption{For L$2_1$ partially ordered CoRuVSi (SQS structure), simulated Berry curvature (-$\Omega_{z}(\bf k)$) (a) along the high symmetry paths and (b) in the k$_x$-k$_y$ plane at E$_F$. Black solid lines show intersections of the Fermi surface with this plane. Fermi surfaces attributed to (c) band 25 and (d) band 26, illustrate the emerging electron and hole pockets again confirming the semimetallic feature. (e) widths of various bands, illustrating the band-crossing at/near $E_F$.}
\label{fig:13}
\end{figure}

The simulated AHE for ordered CoRuVSi (type I configuration) is $\vert\sigma_{int}^{AHE}\vert$=102 $S/cm$, a reasonably high value. The calculated $\vert\sigma_{int}^{AHE}\vert$ is almost double as compared to the experimentally measured value (45 S/cm).
It is important to note that the simulated net magnetization ($\simeq$2 $\mu_B/f.u. $) of the completely ordered phase is quite different as compared to the measured value ($\simeq$ 0.13 $\mu_B/f.u.$) To unveil the possible reason for these discrepancies between theory and experiment, we have simulated the band structure including Berry curvature, anomalous Hall conductivity and Fermi surface of the L$2_1$-disordered phase (50\% disorder between tetrahedral site atoms Co and Ru, as confirmed by XRD-refinement) of CoRuVSi using a 64 atom SQS cell.
This disordered structure gives a reduced magnetization of 0.29 $\mu_B/f.u. $ with a nearly compensated ferrimagnetic structure (see Table \ref{tab:theory-CRVS} for the optimized lattice parameter and the atom projected and total moments). This value of net moment agrees fairly well with our experimental finding. Interestingly, in the L2$_1$ disordered structure, CoRuVSi shows a spin-semimetal behavior (see Fig. \ref{fig:soc}(a)). This is due to a small overlap between the conduction and valence bands (CB and VB) close to the E$_F$, which in turn is tunable by the influence of impurity/disorder or external field, and hence plays a crucial role in the overall electronic structure of the material. To crosscheck the effect of SOC on SSM behavior, we have also simulated the electronic structure including SOC effect. This is shown in Fig. \ref{fig:soc}(b,c).
A close inspection of this band structure in the disordered phase reveals the existence of a linear band crossing at/around -0.4 eV below E$_F$, supported by a band inversion near the $X$ point (see the inset of Fig. \ref{fig:soc}(c)). The simulated Berry curvature, Fermi surface (for band \#25, 26) and the band positions for the L2$_1$ disordered phase of CoRuVSi are shown in Fig. \ref{fig:13}.
Interestingly, we also found remarkable agreement between the calculated AHE ($\vert\sigma_{int}^{AHE}\vert$=52.2 $S/cm$) of the L2$_1$ partially disordered phase and the corresponding experimental value ($\vert\sigma_{int}^{AHE}\vert$=45 $S/cm$). The semimetallic bands, which give rise to the topological non-trivial features are found to be robust against the disorder as indicated by the Berry curvature and Fermi surface calculations.

\begin{table}[t]
\centering
\caption{For L$2_1$ partially ordered CoRuVSi (SQS structure), optimized lattice parameter ($a_0$),  total and atom-projected average moments ($ \mu_B$).}
\begin{tabular}{l  c c c c }
\hline \hline
$a_0$ (\AA)\ \ \ \ \ \ \ \   &  $m^{\mathrm{Co}}$ \ \ \ \ \  \ \ \ &  $m^{\mathrm{Ru}}$ \ \  \ \ \ \ \ \  &  $m^{\mathrm{V}}$   \ \ \ \ \ \ \ \ &  $m^{\mathrm{Total}}$  \\ \hline
5.85\ \ \ \ \ \ \ \ \ \   & 0.29\ \ \ \ \ \ \ \ \ \	& 	 -0.10\ \ \ \ \ \ \  \ \ \ 	& 	 0.10\ \ \ \ \ \ \ \ \ \	& 0.29	 \\
\hline \hline
\end{tabular}
\label{tab:theory-CRVS}
\end{table}

\section{Summary and Conclusion}
In summary, we report the identification of a new member, namely CoRuVSi, to the quantum material class namely {\it spin semi-metals} which can be quite promising for future spintronic and thermoelectric applications.   Using a combined theoretical and experimental study, we have investigated the structural, magnetic, transport, and electronic properties of CoRuVSi. It crystallizes in the cubic structure (space group $F\bar{4}3m$) with a partial L2$_1$-type disorder in the tetrahedral site atoms Co/Ru, as confirmed from our XRD measurement. The magnetization data indicate a weak ferrimagnetic ordering at low T, with a very small moment $\sim$ 0.12  $\mu_B$/f.u. caused by the anti-site disorder. Resistivity results provide a strong evidence of semimetallic nature dominated by two-band conduction, while low-T magnetoresistance data indicate the non-saturating, linear positive magnetoresistance. The latter hints toward the small-gap electronic structure near the Fermi level, indirectly supporting the prediction of semimetallic nature. Specific heat data confirm a low value of density of states at/near E$_F$, supporting our theoretical findings about the semimetallic nature. PCAR measurements reveal a high spin polarization of $\sim$ 50\%. CoRuVSi also shows a high thermopower value of $0.7$ $m Watt/ m-K^{2}$ at room temperature, rendering it as a promising thermoelectric material as well. {\it Ab-initio} simulation of CoRuVSi with L2$_1$ disorder reveals a spin semimetal feature with nearly compensated ferrimagnetic configuration having a small net magnetization, as observed experimentally.  Interestingly, the band structure hosts a linear band crossing at $\sim$-0.4 eV below the Fermi level, along with a band inversion, confirming the topological non-trivial nature of CoRuVSi. \textcolor{black}{This was further assessed from the simulated Berry curvature, anomalous Hall conductivity and Fermi surface.} The simulated anomalous Hall conductivity for L2$_1$ partially ordered CoRuVSi is 52 S/cm which agrees fairly well with experimentally measured value of 45 S/cm. The coexistence of many promising features in a single material is rare and hence it opens up new opportunities to search for other novel materials with multifunctional properties.


{\it \bf Acknowledgments:} JN acknowledges the financial help provided by IIT Bombay. JN also thanks Mr. Vinay Kaushik UGC-DAE-CSR Indore for setting up HC measurements. The authors thank Dr. Durgesh Singh for setting up TEP measurements.  KGS acknowledges the funding from the Indo-Russian Project-TPN- 64868. A.A. acknowledges DST-SERB (Grant No. CRG/2019/002050) for funding to
support this research.


\bibliographystyle{apsrev4-2}
\bibliography{references}

\end{document}